\documentclass[aps,prb,twocolumn,superscriptaddress]{revtex4}
\usepackage{graphicx}
\usepackage{rotating}
\usepackage{float}
\usepackage[colorlinks,citecolor=blue,urlcolor=blue, linkcolor=blue]{hyperref}
\usepackage{epstopdf}   
\graphicspath{{figures//}}

\begin{document} 
  
\title{Electronic effects in high-energy radiation damage in tungsten}
\author{E. Zarkadoula}
\affiliation{Materials Science \& Technology Division, Oak Ridge National Laboratory, Oak Ridge, TN 37831, USA}
\affiliation{Queen Mary University of London, Mile End Road, London, E1 4NS, UK}
\affiliation{SEPnet, Queen Mary University of London, Mile End Road, London E1 4NS, UK}
 
\author{D. M. Duffy}
\affiliation{London Centre for Nanotechnology, Department of Physics and Astronomy, University College London, Gower Street, London, WC1E 6BT, UK}

\author{K. Nordlund} 
\affiliation{University of Helsinki, P.O. Box 43, FIN-00014 Helsinki, Finland}

\author{M. A. Seaton}
\affiliation{Scientific Computing Department, STFC Daresbury Laboratory, Keckwick Lane, Daresbury, Warrington, Cheshire, WA4 4AD, UK} 
 
\author{I. T. Todorov}
\affiliation{Scientific Computing Department, STFC Daresbury Laboratory, Keckwick Lane, Daresbury, Warrington, Cheshire, WA4 4AD, UK} 

\author{W. J. Weber} 
\affiliation{Materials Science \& Technology Division, Oak Ridge National Laboratory, Oak Ridge, TN 37831, USA}
\affiliation{Department of Materials Science \& Engineering, University of Tennessee, Knoxville, TN 37996, USA}

\author{K. Trachenko}
\affiliation{Queen Mary University of London, Mile End Road, London, E1 4NS, UK}
\affiliation{SEPnet, Queen Mary University of London, Mile End Road, London E1 4NS, UK}

\begin{abstract}
Although the effects of the electronic excitations during high-energy radiation damage processes are not currently understood, it is shown that their role in the interaction of radiation with matter is important. We perform molecular dynamics simulations of high-energy collision cascades in bcc-tungsten using the coupled two-temperature molecular dynamics (2T-MD) model that incorporates both the effects of electronic stopping and electron-phonon interaction. We compare the combination of these effects on the induced damage with only the effect of electronic stopping, and conclude in several novel insights. In the 2T-MD model, the electron-phonon coupling results in less damage production in the molten region and in faster relaxation of the damage at short times. These two effects lead to significantly smaller amount of the final damage at longer times.
\end{abstract}


\maketitle
 
\section{Introduction}  

Tungsten and tungsten alloys are candidate materials for the divertor armor in fusion reactors due to their high melting point, high thermal conductivity, reduced deuterium retention and low sputtering erosion \cite{c1,c2,c3,hoen}. Tungsten has been recently used as divertor armor material in recent fusion experiments \cite{jet2}, where it is subject to intense irradiation. The 14 MeV neutrons produced in deuterium-tritium fusion can transfer up to 300 keV energy to a tungsten atom, which will induce the creation of a collision cascade in the system, resulting in the creation of point defects and defect clusters. The evolution of the defect structures over time results in microstructure changes that can induce embrittlement and reduce the thermal conductivity of the material. Importantly, the Joint European Torus (JET) has recently produced tungsten samples irradiated in fusion reactions sustained for several minutes \cite{jet2} providing unique samples irradiated in real fusion reactor conditions. Hence, research into high energy radiation damage in tungsten is particularly timely and of great importance.

Previous studies have investigated radiation effects in tungsten due to lower energy primary knock-on atoms (PKAs) up to some tens of keVs \cite{fu1,troev,fikar1,park1,finnis-w}. A recent study of Sand et al \cite{dudarev_w} studied collision cascades due to 150 keV PKAs at 0 K, taking into account the electronic stopping mechanism which is important at this high energy. In this work we extend the PKA energies to the maximum value expected from 14 MeV neutrons (300 keV) and we investigate the temperature effects. In addition, we include the effects of electron-phonon coupling in our simulations. At high PKA energies a high proportion of energy is lost by interaction with electrons and, as a result, the electronic temperature will rise and affect the defect creation and evolution. Particles moving with velocities that correspond to this high energy lose an important part of their energy due to their interaction with the electrons, hence it is important to include the electronic effects in the simulations. We simulate high--energy radiation damage effects in tungsten at room temperature and at 800 K, which is a typical first wall temperature in fusion reactors.  

In this study we investigate both the effects of electron-phonon coupling and high temperatures on the defect production following 300 keV radiation events in W. We perform two types of radiation simulations. In the first set, which we refer to as friction cascades, we include the electronic stopping using the standard procedure of including a friction term in the equation of motion. In the second set we employ the 2T-MD method for the cascades. In this method the electronic stopping is again included with a friction term, however the energy lost by electronic stopping is deposited in the electronic system . The evolution of the electronic energy  is calculated using a finite difference solution of the heat diffusion equation and coupled to the molecular dynamics (MD) simulation via a Langevin thermostat. This method  transports the excess energy from the simulation cell via electronic thermal conduction and, in addition to including electron-phonon coupling effects, it gives a realistic description of heat transport in the cascade. Both sets of simulations for carried out at 300 K and 800 K to investigate the effect a high temperature environment has on defect production. We compare the peak defect numbers,  the residual defect numbers and the cluster statistics for the four sets of simulations and discuss the reasons for the observed differences in the results.

\section{Methods}

We simulate six randomly chosen directions of the PKA in systems consisting of about 30 million atoms at 300 K and at 800 K. The atoms contained in the boundary of the MD box, in a layer of about 10 \AA\ thickness, are connected to a thermostat so that their velocities are rescaled according to a Maxwell-Boltzmann distribution to correspond to the target temperature. A variable timestep is used to describe the atomic motion throughout the cascade development and relaxation. In the 2T-MD simulations, the atomic and the electronic systems are initially in equilibrium - therefore the temperature of the electronic system at the start of the simulation is equal to the lattice temperature (300 K and 800 K for the two sets of simulations).  The electronic temperature increases during the early stages of the simulation as the energy lost by electronic stopping is input to the electronic system. We are using a modified embedded-atom potential for tungsten \cite{w-pot} joined to the short range repulsive ZBL potential using a polynomial spline, fitted to reproduce the experimentally obtained defect threshold energies. We ran the simulations on up to 32,000 parallel processors of the HECToR National Supercomputing Service.

The electronic effects are included in the simulations according to the Duffy and Rutherford model \cite{duffy0,duffy1} as described in \cite{eza_Fe2}.  For the friction cascades only the friction term that corresponds to the electronic stopping is switched on. The electronic stopping friction coefficient was calculated using SRIM tables \cite{srim} to a value of $1.1$ ps$^{-1}$ and is applied to atoms that move with velocity larger than a value that corresponds to double the cohesive energy of the system \cite{coh2}.  The friction coefficient due to electron–-ion interactions corresponds to coupling parameter value of \mbox{$g_{\mathrm{p}} = 7 \times 10^{17}$ W m $^{-3}$ K$^{-1}$} \cite{w_g}. 
The heat capacity $C_e$ given for a range of electron temperatures can be found in \cite{epc} and was obtained through ab initio calculations as described in \cite{zhigi2}. The values used for the thermal conductivity are $\kappa_{\mathrm{e}} = 174$ W m$^{-1}$ K$^{-1}$ and $125$ W m$^{-1}$ K$^{-1}$ at room temperature and at 800 K \cite{handbook,y.g.li} respectively. Due to large uncertainty we assume no lattice temperature dependence in $\kappa_{\mathrm{e}}$ and a constant value of $g_{\mathrm{p}}$.  The value of $\kappa_{\mathrm{e}}$ is accurate for 300 K and 800 K, but it decreases with lattice temperature; therefore it is likely to be overestimated in the thermal spike region. Higher thermal conductivity will also increase the temperature of the thermal spike. The value of $g_{\mathrm{p}}$ is within the experimentally obtained range of $5-10 \times 10^{17}$ W m $^{-3}$ K$^{-1}$ \cite{w_g}, however recent ab initio calculations \cite{daraszewicz_w} suggest that this value might be overestimated. Higher value of g will have the effect of increasing the rate of energy transfer from ions to electrons and, therefore, cooling the thermal spike faster.

The e-ph coupling process $g_{\mathrm{p}}$ is activated at 0.3 ps of simulation time, as the lattice temperature is ill-defined before this. Until this time of the simulation only the electron stopping mechanism is active. This approximate value was computed by looking at the convergence of kinetic and potential energies (i.e. thermalization) in the friction cascades. To solve the electronic thermal diffusion equation, the system volume is divided into smaller voxels. Each voxel has an associated electronic temperature and its atomistic temperature is calculated from the kinetic energy of atoms within it. Further voxels are included beyond the system volume to transport energy away from the collision cascades \cite{duffy1}.

We use the sphere criterion for defect identification \cite{Nor97f,dlpoly_manual}. $N_{\mathrm{disp}}$ accounts for the total displacements
introduced in the system, i.e., the number of atoms that have moved more than a cut-off distance ($d=0.75$ \AA) from their initial positions. To account for the atoms that recombine to crystalline positions, $N_{\mathrm{def}}$ is introduced. $N_{\mathrm{def}}$ reflects the recovery of structural damage as it corresponds to the sum of interstitials and vacancies.  $d$ should generally be smaller than half of the closest interatomic separation, and is usually chosen not to account for typical thermal fluctuations of $0.2$–-$0.3$ \AA.

\section{Results and Discussion}

\subsection{Electron-phonon coupling effect}

\begin{figure*} 
\begin{centering}
\includegraphics[width=6in]{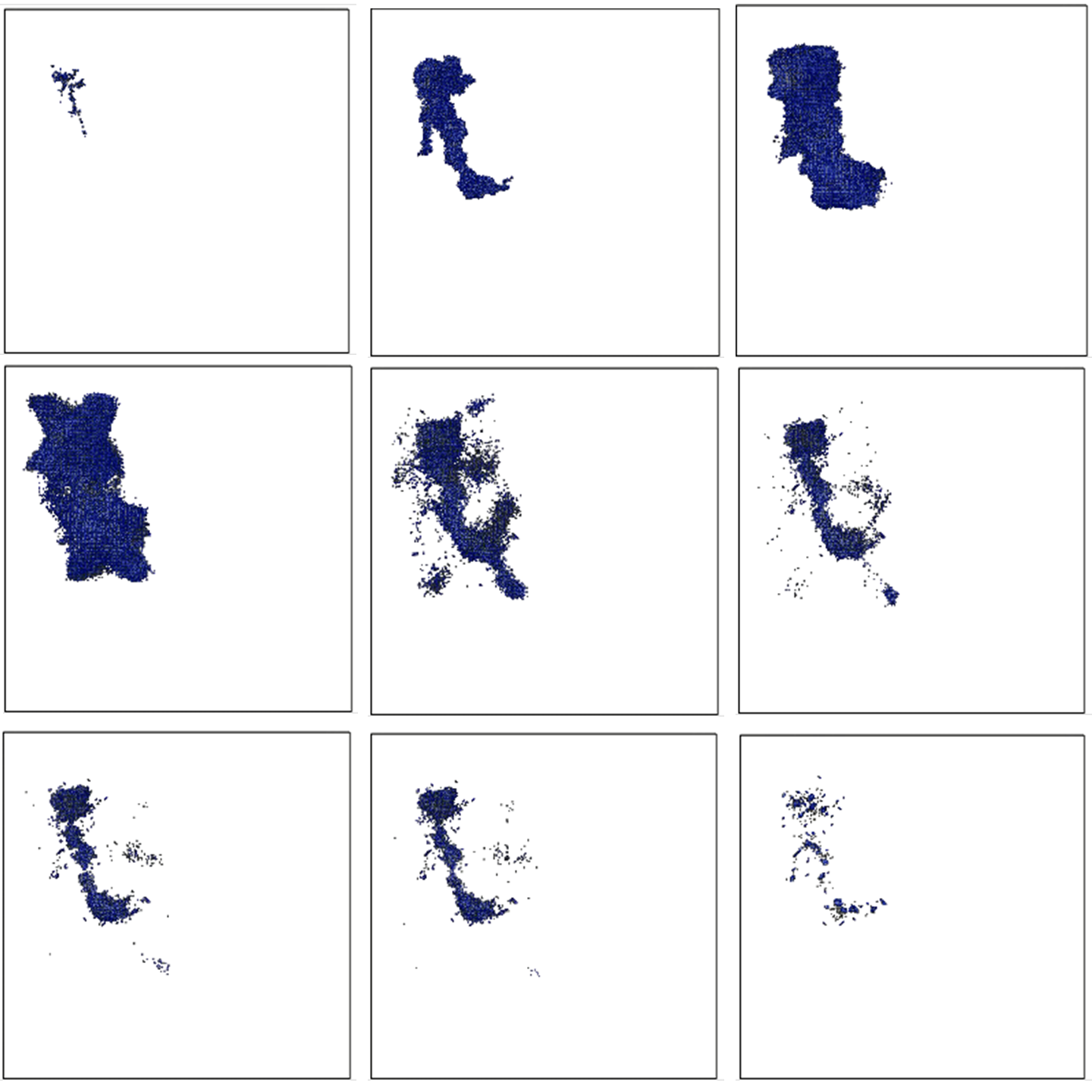}
\end{centering}
\caption{Relaxation of a 300 keV 2T-MD cascade performed at 800 K in a system consisting of about 30 million atoms. The simulation box has about 800 \AA\ length. The simulation time is about 100 ps. Interstitials are shown in blue and vacancies in grey.
}
\label{fig1} 
\end{figure*}

In Fig. \ref{fig1} we show the defect evolution of a typical 300 keV 2T-MD cascade at 800 K, in a system consisting of about 30 million atoms. The simulation box has a length of about 800 \AA\ and the cascade size is about 450 \AA. This figure illustrates continuous morphology of the collision cascade. This absence of sub-cascades is in agreement to our results for higher energy events in iron and zirconia \cite{eza,eza_zro2}.

\begin{figure*}
\begin{centering} 
\includegraphics[width=6.5in]{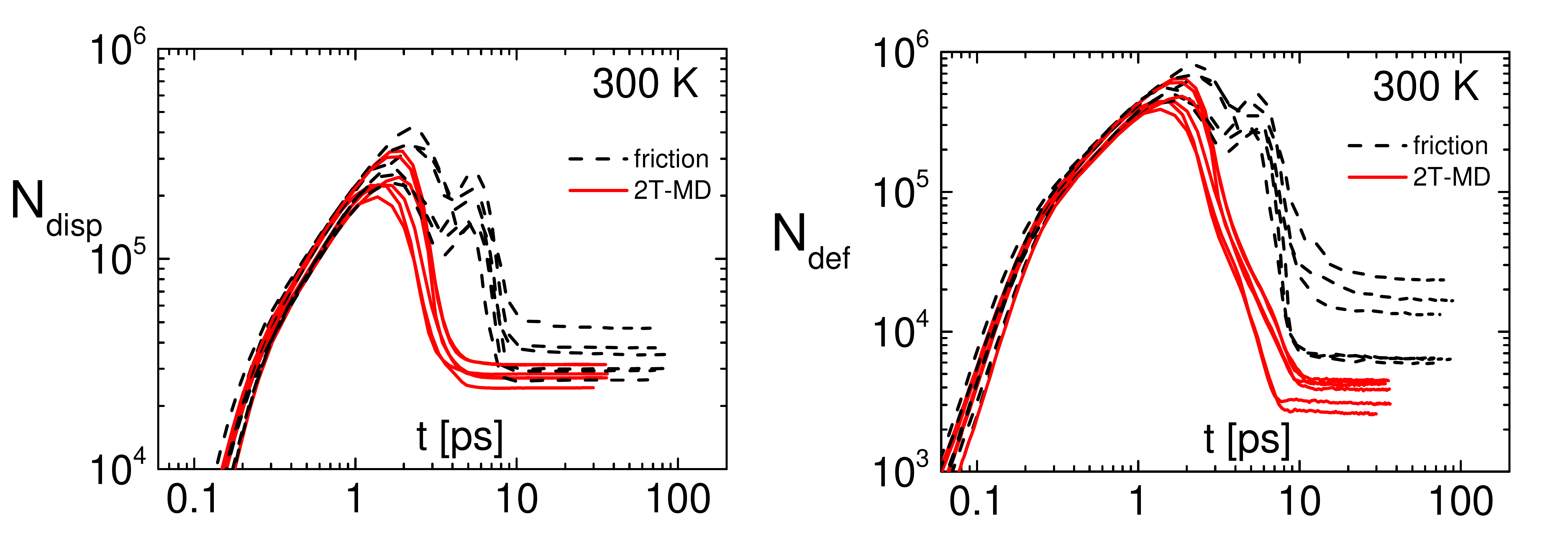}
\end{centering} 
\caption{Number of displaced (left) and defect ($N_{\mathrm{def}}=N_{\mathrm{int}}+N_{\mathrm{vac}}$)(right) atoms for six 300 keV friction cascades (black dotted lines) and six 2T-MD cascades (red solid lines) at 300 K.
}
\label{fig2}
\end{figure*}

\begin{figure*}
\begin{centering}
\includegraphics[width=6.5in]{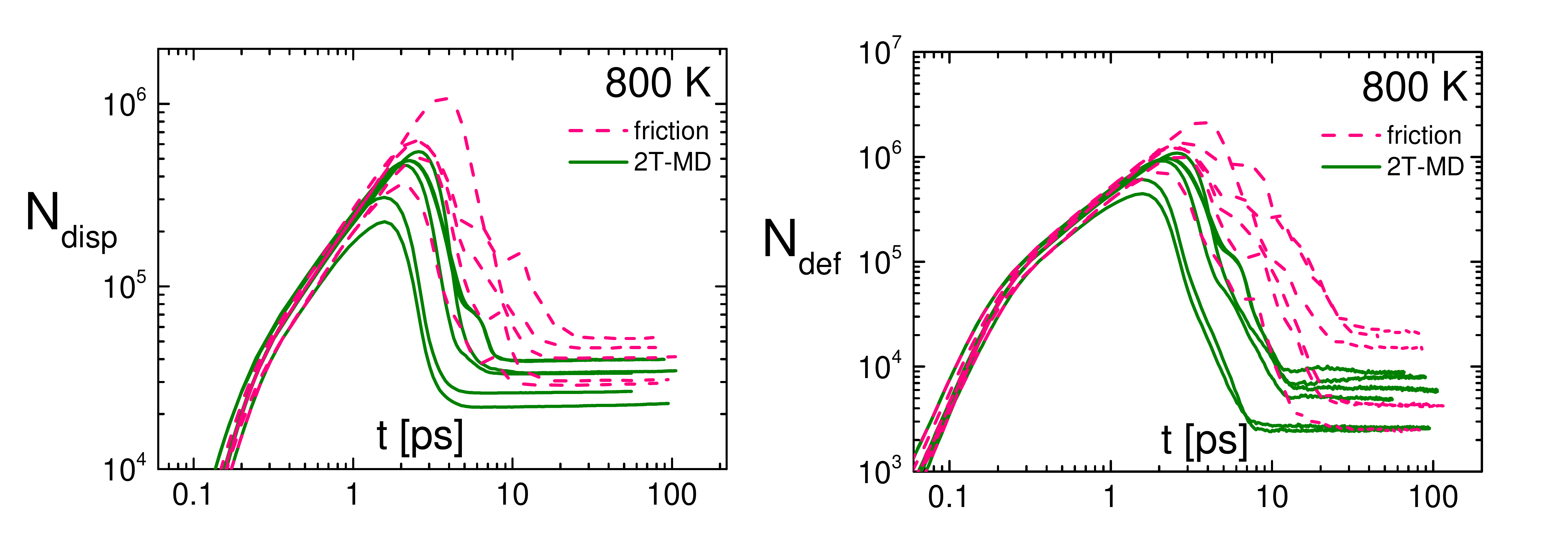}
\end{centering}
\caption{Number of displaced (left) and defect ($N_{\mathrm{def}}=N_{\mathrm{int}}+N_{\mathrm{vac}}$)(right) atoms for six 300 keV friction cascades (pink dotted lines) and six 2T-MD cascades (green solid lines) at 800 K.
}
\label{fig3}
\end{figure*}

Damage creation and recovery for 300 K friction and 2T-MD cascades and 800 K friction and 2T-MD cascades is illustrated in Fig. \ref{fig2} and \ref{fig3}, respectively. We will use $N_{\mathrm{disp}}^p$ to refer to the peak of displaced atoms and with $N_\mathrm{def}^p$ to the peak of the defect atoms, and $N_{\mathrm{disp}}^l$ and $N_\mathrm{def}^l$ to the number of displaced and defect atoms at long simulation times (the flat lines in Fig. \ref{fig2}-\ref{fig3}. $\tau_{\mathrm{disp}}$ and $\tau_{\mathrm{def}}$ are the relaxation times that correspond to the width of $N_{\mathrm{disp}}^p$ (elastic deformation) and to the width of $N_\mathrm{def}^p$ (dynamic annealing) respectively.

In Figures \ref{fig2} and \ref{fig3} we see $N_{\mathrm{disp}}$ and $N_{\mathrm{def}}$ as a function of time for friction and 2T-MD cascades performed at 300 K and at 800 K respectively. In agreement with our findings for iron, the relaxation times $\tau_{\mathrm{disp}}$ and $\tau_{\mathrm{def}}$ are smaller than the relaxation times for the friction cascades when the 2T-MD model is applied. For the friction cascades $\tau_{\mathrm{disp}}$ and $\tau_{\mathrm{def}}$ are about 10-15 ps, and about 5 ps and 10 ps for the 2T-MD cascades.

The narrower peak of the 2T-MD cascades shows less damage creation when the e-ph coupling is applied: the e-ph coupling in the 2T-MD model removes energy from the thermal spike for both temperatures resulting in less defect creation and, importantly, less damage at the end of the simulation time. On average, the fraction of $N_\mathrm{def}^l(\mathrm{friction})$/$N_\mathrm{def}^l$(2T-MD) for the cascades at 300 K at the end of the simulation time is about $3$. For the cascades performed at 800 K, the fraction of $N_\mathrm{def}^l(\mathrm{friction})$/$N_\mathrm{def}^l$(2T-MD) at the end of the simulation time is $2$. The second peaks in the friction model runs are due to reflection of the excess energy from the boundaries of the MD cell. The MD cell boundary thermostat cannot absorb the excess energy with perfect efficiency and it reflects it towards the molten region. In the 2T-MD model, where energy is transported out of the simulation cell by the electrons, we do not see reflection of the excess energy from the boundary thermostat. 

\begin{figure}
\begin{centering}
\includegraphics[width=\columnwidth]{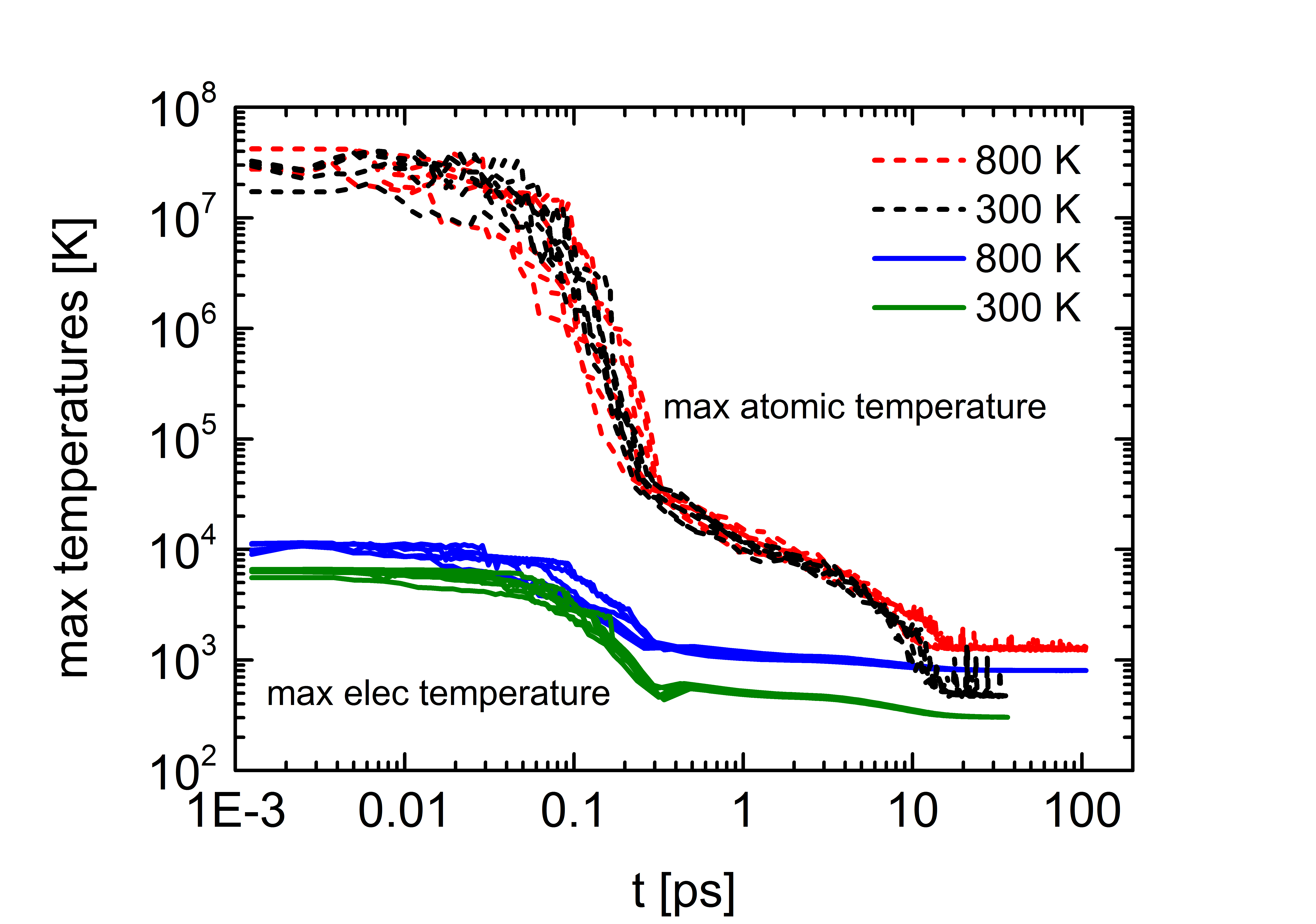} 
\end{centering}
\caption{Maximum electronic and atomic temperatures for 300 keV 2TMD cascade simulations at 300 K and at 800 K, for six events at each temperature.
The ill-defined lattice temperature reaches past $10^7$ K initially. After 0.3 ps, which is the thermalization time, electronic energy is fed back
to the lattice and the ionic temperature starts dropping below $10^4$ K. At around 11 ps the electron–ion temperatures are equilibrated.
}
\label{fig4}
\end{figure}

Figure \ref{fig4} shows the maximum electronic and atomic temperatures for 300 keV 2TMD cascades at 300 K and 800 K. The maximum electronic temperature is the highest per-voxel value to be found in the grid. For all simulations the atomic temperature is higher than the electronic temperature, meaning that the electronic system acts as a heat sink. This is in agreement with 2T-MD cascades in iron \cite{duffy0, eza_Fe2}. The heat transfer relaxation time, as read from these plots, is about 11 ps.

The small difference in the produced damage at the peak and the significant difference in the remaining damage in the system are in contrast to our findings for iron \cite{eza_Fe2}. For iron we found that the 2T-MD model results in reduced damage at the peak (the peaks shown in the plots of Fig. \ref{fig2} and \ref{fig3}) and in small difference of remaining defects. This difference could be due to different values of parameters in the 2T-MD model, such as the e-ph coupling parameter or the thermal conductivity.

\subsection{Temperature effect}
 
\begin{figure*}
\begin{centering}
\includegraphics[width=6.5in]{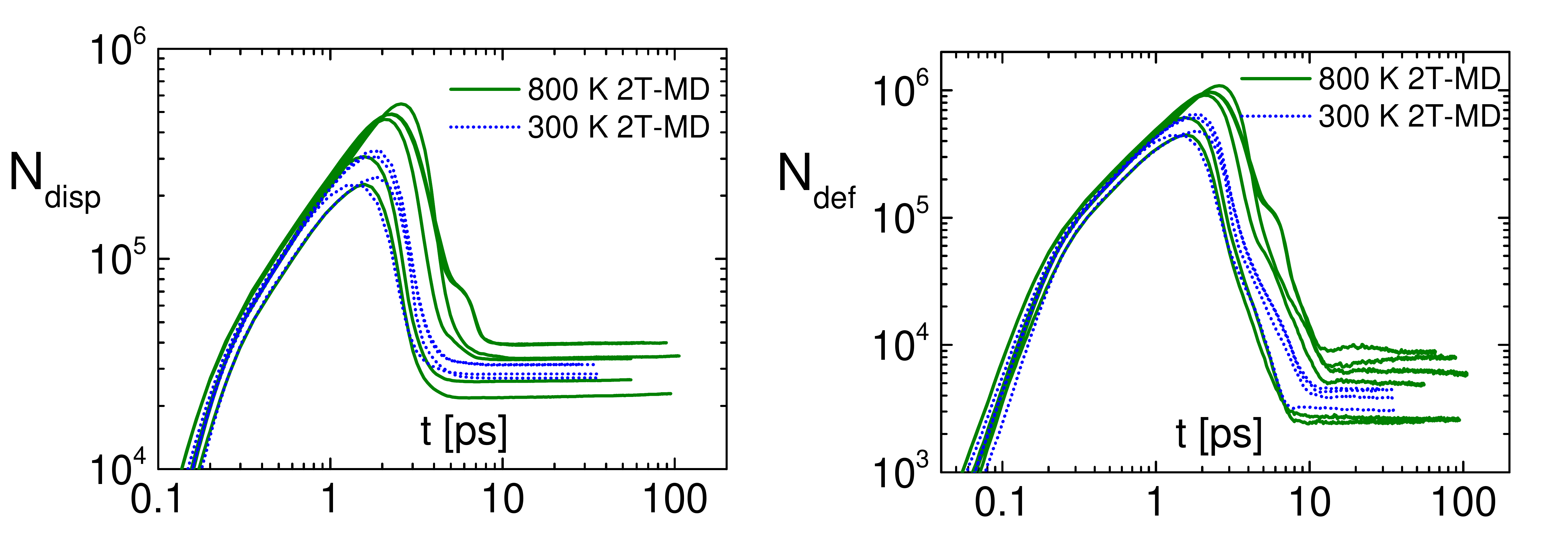}
\end{centering}
\caption{Number of displaced (left) and defect ($N_{\mathrm{def}}=N_{\mathrm{int}}+N_{\mathrm{vac}}$) (right) atoms for six 300 keV 2T-MD cascades at 300 K (blue dotted line) and at 800 K (green solid line).
}
\label{fig5}
\end{figure*}

Figure \ref{fig5} illustrates $N_{\mathrm{disp}}$ and $N_{\mathrm{def}}$ for six collision cascades at 300 K (blue dotted line) and at 800 K (green solid line). Here $N_\mathrm{disp}^p$ and $N_\mathrm{def}^p$ are larger for the 800 K collision cascades than for the 300 K, which means that there is on average a larger number of displacements and damage creation in the thermal spike for the cascades performed at higher temperature. Consequently, the resulting displacements $N_\mathrm{disp}^l$ and damage $N_\mathrm{def}^l$, as seen in the flat lines at long simulation time,  are larger for the cascades performed at 800 K. This is in contrast to results of previous temperature dependent simulations in materials different than tungsten, such as iron \cite{stoller_temp1,stoller_temp2} and gold \cite{Nor96d}. However, recent modeling results of lower energy displacement cascades in tungsten \cite{nandipati-i,nandipati-ii} show no systematic recombination of self interstitial atoms (SIAs) with temperature: low energy cascades show an increase of surviving Frenkel pairs (FP) from 1025 K to 2050 K, and cascades of 100 keV PKA show similar numbers of surviving FP at 300 K, 1025 K and 2050 K. In our simulations, as shown in Fig. \ref{fig5} we have a larger molten region for the cascades performed at 800 K (wider peaks for the 800 K 2T-MD cascades), which survives annealing, resulting in higher number of defects. On average, the fraction of $N_\mathrm{def}^l(\mathrm{800})$/$N_\mathrm{def}^l(\mathrm{300})$ at the end of the simulation time is about $1.5$. Similar simulations without inclusion of the electronic effects result in a defect ratio $N_\mathrm{def}^l(\mathrm{300})$/$N_\mathrm{def}^l(\mathrm{800})$ of 2.3, which supports the fact that the results discussed here are due to the electronic effects.

\subsection{Defects at local level}

Defect analysis at the local level is summarized in Table \ref{tabul1}, where we give statistics for the defect clusters for the cascades of this paper. Self-interstitial atoms (SIAs) and vacancies are defined as belonging to the same cluster if they are within two nearest-neighbor distances. As discussed above, compared to friction only cascades, the 2T-MD cascades result in less damage. When the friction alone is applied, the defect production does not depend on the temperature. However, in the 2T-MD cascades, higher temperature results in increased damage. As shown in the table, we found that most of the vacancies and most of the interstitials are organized in clusters. The fraction of vacancies and interstitials in clusters of size 4 or more does not significantly change with temperature or by including the e-ph interaction. We find that the number of vacancy and SIA clusters decreases with increasing temperature and decreases when the full 2T-MD model is applied. The latter is in agreement with results of the 2T-MD model in 200 keV cascades in bcc-iron \cite{eza_Fe2}, where the 2T-MD model results in smaller size of interstitial clusters for both simulated temperatures. 

We analyzed 10 of the largest interstitial clusters from the 2T-MD cascades at 300 K and 10 from 2T-MD cascades at 800 K. The count of the interstitials (as opposed to the net defect count given in Table \ref{tabul1}) contained in these clusters is $438$-$1754$ and $391$-$5920$ respectively. We found almost half of them to be dislocation clusters and loops with $<$111$>$ Burgers vector. The rest have either $<$100$>$ Burgers vector or they consist of clusters with $<$111$>$ and $<$100$>$ Burgers vectors. These results are with agreement with recent experimental results of irradiation of tungsten at 500 C \cite{yi} and also with agreement of recent modeling results of tungsten irradiation with energies up to 100 keV at 0 K \cite{dudarev_w}. We analyzed the 10 largest vacancy clusters (consisting of $125$-$527$ vacancies) from the 2T-MD cascades at 300 K and the 10 largest vacancy clusters (consisting of $176$-$2878$ vacancies) from the 2T-MD cascades at 800 K and found no vacancy dislocation loops in agreement with the recent MD results of Sand et al. \cite{dudarev_w} for the same potential.

\begin{figure}
\begin{centering} 
\includegraphics[width=\columnwidth]{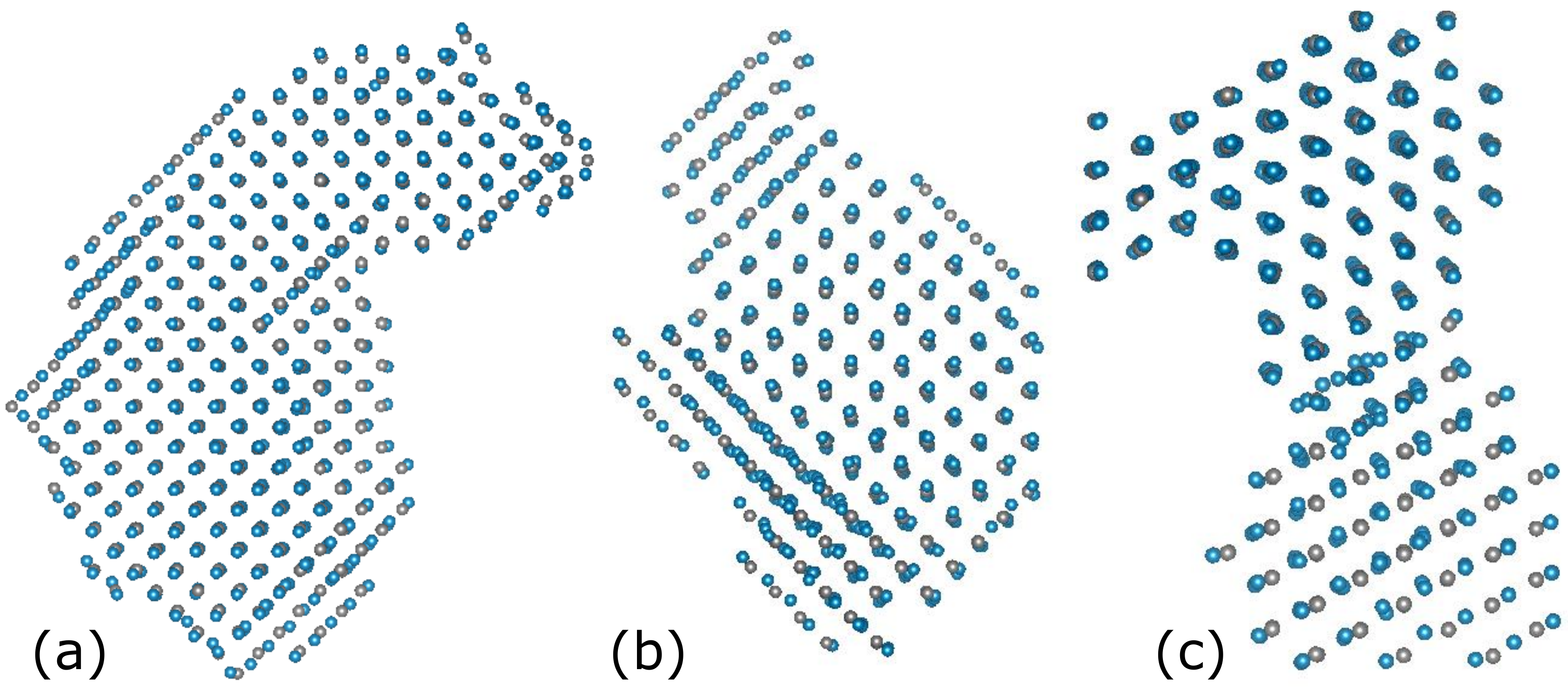}
\end{centering}
\caption{(a) and (b) Dislocation-like clusters and in the $<$100$>$  in typical 2T-MD 300keV cascades performed at 300 K.  (c) (111) projection of a mixed ($<$111$>$ and $<$100$>$) cluster in a 2T-MD 300keV cascades performed at 300 K. Interstitials are shown in blue and vacancies in grey. 947 interstitial atoms and 807 vacancies are shown in (a), 534 interstitial atoms and 447 vacancies in (b), and 543 interstitials and 474 vacancies in (c). Cluster sizes are 3-5 nm.
}
\label{fig6}
\end{figure}

\begin{figure}
\begin{centering} 
\includegraphics[width=\columnwidth]{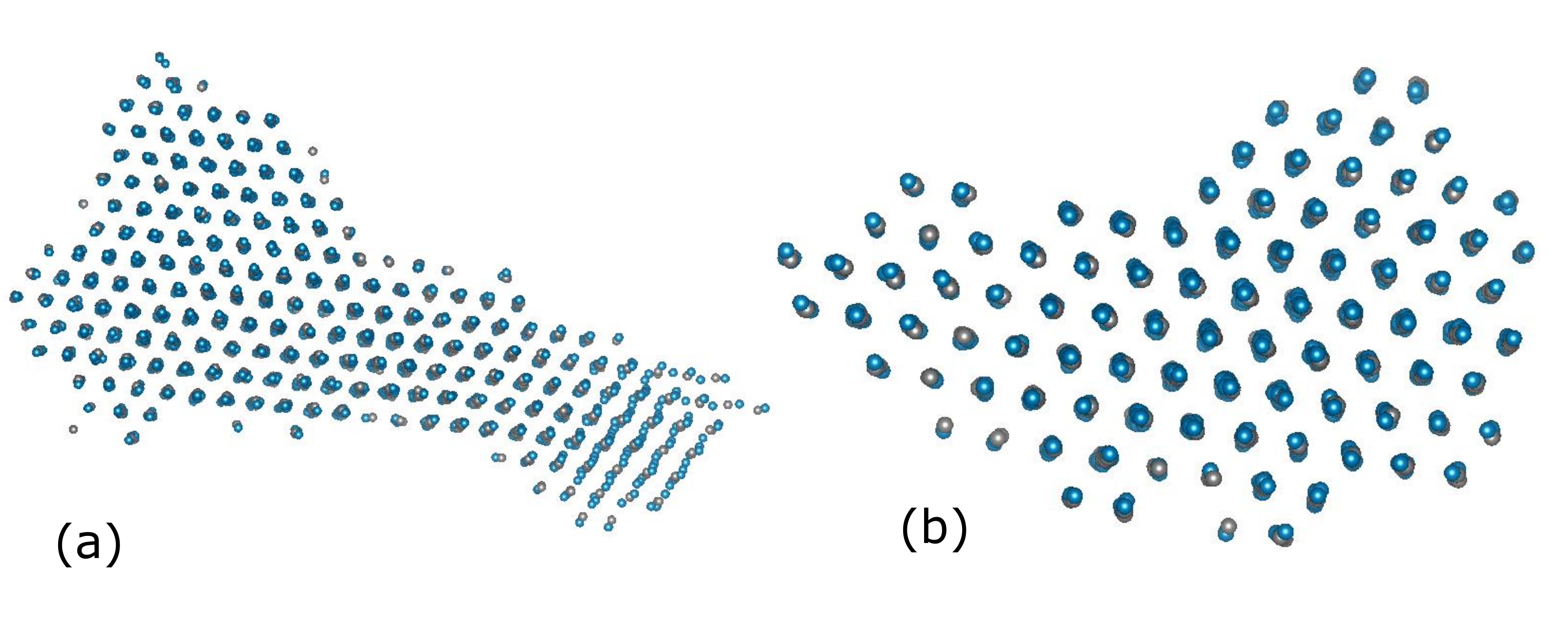}
\end{centering}
\caption{Dislocation-like clusters and loops in two typical 2T-MD 800keV cascades performed at 800 K. (a) (111) projection of a mixed cluster ($<$111$>$ and $<$100$>$) (b) typical $<$111$>$ cluster. Interstitials are shown in blue and vacancies in grey. The number of interstitial atoms and vacancies shown are respectively 2483 and 2319 for (a), and 945 and 863 for (b). The cluster sizes are about 5-7 nm. 
}
\label{fig7}
\end{figure}

In Fig.\ref{fig6} we show typical clusters found in 2T-MD cascades at 300 K. The $<$100$>$ cluster in (a) is the largest interstitial cluster in 2T-MD cascades at 300 K with net defect number (difference between the number of SIAs and the number of vacancies) 140. In (b) we show a $<$100$>$ cluster and in (c) a combination of a $<$111$>$ cluster and a $<$100$>$ cluster. The $<$111$>$ projection of the largest SIA cluster for 2T-MD cascades at 800 K (net defect number 164) is shown in Fig. \ref{fig7} (a). This is composed of a $<$111$>$ cluster and a smaller $<$100$>$ cluster. Fig. \ref{fig7} (b) shows a typical $<$111$>$ cluster found in a 2T-MD cascade at 800 K. The sizes of the clusters analyzed here (3-8 nm in diameter) agree with the findings of Yi et al. \cite{yi}.

\begin{table*}[tb] 
\begin{center}
\setlength{\tabcolsep}{0.3cm} 
\begin{tabular}{lcccc}
{\bf Model}  & {\bf Friction}  & {\bf  2T-MD}  & {\bf  Friction}  & {\bf 2T-MD}  \\
            & {\bf 300K} & {\bf 300 K}    & {\bf 800 K} & {\bf 800 K}   \\ [2ex]
            \hline
Number of Frenkel Pairs              							  & 5500 (1100) & 1900 (200) & 5500 (1700) & 2700 (500)    \\[1ex]
Fraction of vacancies in clusters of size more than 3 & 0.88 (0.02)    & 0.75 (0.02)    & 0.89 (0.02)     & 0.82 (0.03)    \\[1ex]
Fraction of SIAs in clusters of size more than 3      & 0.98 (0.01)    & 0.93 (0.18)    & 0.97 (0.01)     & 0.96 (0.01)    \\[1ex]
Number of vacancy clusters  						  & 110 (8)   & 100 (3)    & 90 (12)     & 70 (7)          \\[1ex]
Number of SIA clusters   						      & 50 (8)     & 80 (5)     & 40 (4)     & 50 (6)           \\[1ex]
Largest SIA~cluster       	 					      & 374         & 140        & 317         & 164           \\[1ex]
\end{tabular} 
\end{center}
\caption{The number of Frenkel pairs,calculated using the sphere criterion, and defect distribution statistics for 300 keV  friction and 2T-MD cascade simulations in tungsten at 300 K and 800 K. The standard error of the mean is shown in the brackets calculated over six events. The largest clusters that we found in each set of six simulations are presented in the last row and are determined by net defect count (difference between the number of SIAs and the number of vacancies).}
\label{tabul1}
\end{table*}

\section{Conclusions}

In summary, we presented the results of high energy radiation damage in tungsten in 300 K and 800 K with full account of the electronic effects. The significant difference in the residual damage demonstrates the need to accurately simulate high-energy collision cascades by including in the simulations a local model to describe the electron--phonon coupling. 
The diversity of the results of the 2T-MD model in iron, where the e-ph coupling results in similar residual damage, and tungsten, where the e-ph coupling leads to a smaller amount of damage, demonstrates the importance of taking into account the electronic effects at these high energies and emphasizes the need to avoid extrapolations of results obtained for one material to others, even in a case like tungsten, which has the bcc structure like iron. 

Our findings show that implementing a model to account for the electron-phonon coupling in high-energy collision cascades can significantly affect the resulting damage in the system quantitatively. Additionally, the electronic effects can significantly affect the fraction of damage in clusters, resulting in different defect structures and different cluster sizes. This highlights that radiation damage in metals requires not only an accurate interatomic potential, but importantly, the consideration of the e-p coupling in a local way. Further investigation of the electronic effects in solids and their role in damage creation is needed in order to understand fundamental aspects of radiation effects in materials and predict their performance under extreme conditions. Additionally, further investigation of the 2T-MD parameters, and in particular their temperature dependence, will improve the accuracy of 2T-MD simulations.   

\section{Acknowledgments} 
We acknowledge Prof Steven Cowley and Dr Tom Todd from Culham Centre for Fusion Energy, Abingdon, Oxfordshire OX14 3EA, United Kingdom, for useful discussions.

\end{document}